\documentclass[]{spie}
 
\usepackage{amsmath,amsfonts,amssymb}
\usepackage{graphicx}
\usepackage[colorlinks=true, allcolors=blue]{hyperref}

\title{Superefficient long-lived multiresonator quantum memory}

\author[a,b]{Nikolay S. Perminov}
\author[c]{Diana Yu. Tarankova}
\author[a,b,*]{Sergey A. Moiseev}
\affil[a]{Kazan Quantum Center, Kazan National Research Technical University n.a. A.N.Tupolev-KAI, 10 K. Marx, Kazan 420111, Russia}
\affil[b]{Zavoisky Physical-Technical Institute of the Russian Academy of Sciences, 10/7 Sibirsky Tract, Kazan 420029, Russia}
\affil[c]{Institute of Radio-Electronics and Telecommunications, Kazan National Research Technical University n.a. A.N.Tupolev-KAI, 10 K. Marx, Kazan 420111, Russia}

\authorinfo{* s.a.moiseev@kazanqc.org}

\begin{document} 
\maketitle

\begin{abstract}
In this paper, we propose a scheme of a long-lived broadband superefficient multiresonator quantum memory in which a common resonator is connected with an external waveguide and with a system of high-quality miniresonators containing long-lived resonant electron spin ensembles.
The scheme with 4 miniresonators has been analyzed in details and it was shown that it is possible to store an input broadband signal field to the electron spin ensembles with quantum efficiency 99.99\%.
The considered multiresonator system opens the way to elaboration of efficient multiqubit quantum memory devices for superconducting quantum computer.
\end{abstract}


\section{INTRODUCTION}
Development of the quantum memory (QM) as well as an effective light-media quantum interface (QI) is of decisive importance for quantum information technologies \cite{Gu2017,Xiang2013,Devoret2013,Kurizki2015,Hammerer2010}. 
Impressive results have been recently achieved in experimental implementation of the effective optical QMs \cite{Hedges2010,Hosseini2011,Cho2016}.
The proposed approaches triggered active elaboration of the microwave QM which could be a key element for the creation of universal superconducting quantum computer \cite{Grezes2014,Gerasimov2014,Flurin2015,Pfaff2017}. 
In practical implementation of long-lived multiqubit QM, it is required an implementation of sufficiently strong and reversible interaction of light/microwave qubits with many \cite{Roy2017} information carriers, in particular with NV-centers in diamond \cite{Jiang2009} and rare-earth ions in inorganic crystals \cite{Zhong2015}.
However, experimental realization of sufficiently high quantum efficiency remains main problem in these investigations.

\begin{figure}[ht]
\begin{center}
\includegraphics[width=0.48\textwidth]{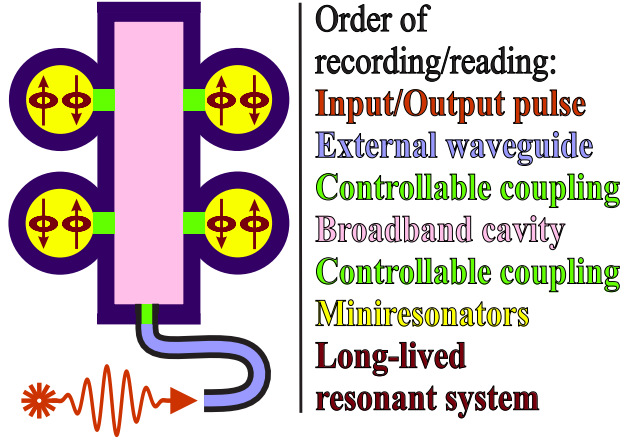}
\end{center}
\caption{
Principal scheme of HMR QM. The color of the text corresponds to the colored elements in the figure on the left.}
\label{scheme}
\end{figure}

\noindent
The rich dynamics and controlling of multiparticle quantum systems \cite{Hartmann2008,Hur2016,Noh2017} with a large number of variable parameters provide ample opportunities for constructing composite quantum circuits with predefined properties, which is required  for creating various elements of a full-fledged multifunctional quantum computer.
In this paper, we propose a scheme for broadband long-lived hybrid multiresonator (HMR) QM (see figure \ref{scheme}).
The memory scheme contains a system of miniresonators placed in a common broadband resonator that is also connected to an external waveguide.
Resonant frequencies of miniresonators constitute a periodic structure of narrow lines.
In turn, each miniresonator contains an ensemble of electron spins (or other resonant particles) whose resonant frequency is tuned to the frequency of miniresonator.
The system of high-quality miniresonators provides an optimal condition for sufficiently strong interaction of the broadband signal field with electron spin ensembles.
Herein, the multiresonator (MR) scheme \cite{Moiseev_2017} playing the role of a broadband QI for an effective reversible transfer of a signal from an external waveguide to the system of electron spins which acts as long-lived carrier of quantum information.

\noindent
Here we consider the case of small number (four) of miniresonators, which greatly facilitates the spectral-topological optimization \cite{Perminov2017} of the parameters of the MR scheme.
The simple model of spectral optimization proposed below was based for the fitting the transfer function (TF) of the studied system to the parameters of TF of ideal QM.
The fitting has been realized at a finite set of spectral points in the selected frequency range.
By using this method we have found the optimal values of all free parameters of the HMR scheme without using large computing resources.
For optimized HMR QM, we used the spectrum of narrow lines and analyzed the spectral dependence of the quantum efficiency, which demonstrated the possibility of achieving the super high efficiency $\sim99.99\%$ in a wide frequency range.
It was shown that the HMR scheme does not impose large restrictions on the parameters of losses in miniresonators and in the common broadband resonator.
We also discuss using of HMR schemes as a highly efficient QI for combining multiple QM devices into a single wideband circuit with a higher-quality and wider spectral profile.

\section{PHYSICAL MODEL}
General theoretical model of the scheme under consideration generalizes the KM model on the photon echo \cite{Moiseev2001} in the optimal resonator \cite{Moiseev2010}, which has been recently extended to systems of several resonators \cite{Moiseev_2017,MoiseevE2017}.
For the HMR system, we analyze the dynamics of $N$ miniresonators and resonant electron spin ensembles by using the quantum optics approach to the description of light in an open cavity \cite{Walls}.
In the framework of this approach we get the system of equations for the spin coherence $s_{n,j}(t)$, the miniresonators field modes $b_n(t)$ and the mode of the common resonator field $a(t)$:
\begin{align}\label{gen_eq}
& \nonumber \left[\partial_{t}+i(\tilde{\Delta}_n+\delta_{n,j})\right]s_{n,j}(t)-f_n^{0}b_n(t)=0,\\
& \nonumber \left[\partial_{t}+\gamma_n+i\Delta_n\right]b_n(t)+\sum_jf_n^0s_{n,j}(t)+[g_n^{0}]^{*}a(t)=0,\\
& \left[\partial_{t}+\gamma_r+\frac{\kappa}{2}\right]a(t)-\sum_ng_n^0b_n(t)=\sqrt{\kappa}a_{in}(t),
\end{align}
where $a_{in}(t)=[2\pi]^{-1/2}\int~d\nu e^{-i\nu t}\phi_\nu$ is the input pulse, $\phi_\nu$ is the spectral profile of the input pulse, for which the normalization condition for the single-photon field is fulfilled $\int d\nu |\phi_\nu|^2=1$, $\nu$ is the frequency counted from the central frequency of the radiation $\omega_0$, $\delta_{n,j} $ is the frequency detuning of the $j$-th spin in the $n$-th spin system (reckoned from its line center) with detuning $\tilde{\Delta}_n $ ($j\in\{1,...,N_n\}$), $\{f_n^0,1/T_2^*\}$ is the field interaction constant in the $n$-th miniresonator with a single electron spin and the inhomogeneous linewidth of the $n$-th spin system, $\{\tilde{\Delta}_n=\Delta(n-\operatorname{sgn}(n)/2),\Delta_n\}$ is the frequency detuning of centers of lines of spin systems and miniresonators, $n\in\{-N/2,…,N/2\}\backslash\{0\}$ (in this paper we consider the case of an even $N$), $\{\gamma_n,\gamma_r\}$ is the decay constant of the field modes of the $n$-th miniresonator and the mode of the common resonator, $\{g_n^0,\kappa\}$ is the coupling constant of the $n$-th miniresonator with a common resonator and the coupling constant of a common resonator with an external waveguide.
Assuming the number of spins in each mini-cavity is large $N_n\gg1$, the value $\delta_{n,j}$ is replaced by the frequency detuning $\delta$, which has a continuous distribution with the Lorentzian line shape $G_n(\delta)=\pi^{-1}(1/T_2^*)[(\delta)^2+(1/T_2^*)^2]^{-1}$.
We also ignored the Langevin forces \cite{Walls} in the equation (\ref{gen_eq}), focusing only on the searching for the quantum efficiency in the HMR scheme studied.

\section{RECORDING/READING STAGE}
For the absorption stage, from the equations (\ref{gen_eq}) we find the output field $a_{out}(t)=[2\pi]^{-1/2}\int dve^{-i\nu t}S(v)\phi_{\nu}$ in terms of the TF $S(\nu)=\tilde{a}_{out}(\nu)/\tilde{a}_{in}(\nu)$ in the following form:
\begin{align}\label{gen_sol}
& S(\nu)=\frac{1-F(\nu)}{1+F(\nu)},\\
& \nonumber F(\nu)=\tilde{\gamma}_r-\frac{2i\nu}{\kappa}+\sum_n
\frac{g_n}{\frac{f_n^2}{1/T_2^*+i(\tilde{\Delta}_n-\nu)}+\gamma_n+i(\Delta_n-\nu)},
\end{align}
where $a_{in,out}(t)=[2\pi]^{-1/2}\int~d\nu e^{-i\nu t} \tilde{a}_{in,out}(\nu)$, $\tilde{\gamma}_r=2\gamma_r/\kappa$, $2|g_n^0|^2/\kappa=g_n$, $N_n(f_n^0)^2=f_n^2$
and also the ratio $a_{out}(t)=\sqrt{\kappa}a(t)-a_{in}(t)$ is taken into account in accordance with the input-output approach \cite{Walls}.
Next we will use dimensionless units for all the parameters in (\ref{gen_sol}), which is equivalent to relating them to a certain unit of frequencies, assuming without loss of generality $\Delta=1$.

\noindent
Since we consider the problem of highly efficient storage for the broadband signal field in the electron spin systems, we will assume that the irreversible losses in each mini-resonator are sufficiently weak $\tilde{\gamma}_r=\gamma_n\sim10^{-1}\div10^{-2}$.
The weak losses do not affect to the optimization process and to the values of optimal physical parameters in HMR scheme.
This argument allows us to simplify the calculations by putting $\tilde{\gamma}_r=\gamma_n=0$ in (\ref{gen_sol}).
For simplicity, we also assume that the common resonator is sufficiently broadband, i.e. $N\Delta/\kappa\ll1$ and $|\nu|/\kappa\ll1$, and the constant $\tilde{\gamma}_r=2\gamma_r/\kappa$ can be small enough, for example, for superconducting resonators, which  is important for practical implementation of the HMR scheme.
If the QM protocol is time reversible, as it is typical for the photon/spin echo schemes \cite{Moiseev2010}, then overall efficiency can be written as $\eta_{total}=(\eta_{recording})^2\cdot\eta_{storage}$ (the storage efficiency $\eta_{storage}$ in the spin system includes the losses associated with the procedure providing time reverse dynamics of electron spins due to excitation with additional external control fields (the topic which goes beyond the scope of this work).
Therefore, we can restrict ourselves to the problem of efficient transfer of the signal field to the system of electron spins (maximization of the recording efficiency $\eta_{recording}$), which is directly related to the optimization of the parameters of the HMR QI: parameters of spin-system, miniresonators, common resonator and its interaction.

\section{OPTIMIZATION OF HMR QM IN A WIDE FREQUENCY BAND}
In general, TF (\ref{gen_sol}) has very complex spectral properties \cite{Swanson2007,Koziel2008,Koziel2011,Tamiazzo2017} (many mathematical aspects of our topic intersect with the fundamental problems of the theory of filters).
However, we show that under the certain conditions, the real system can acquire the spectral properties of TF corresponding to the highly effective sensor \cite{Cheng2014}, broadband filter \cite{Kozakowski2005,Swanson2007,Rosenberg2013}, QM \cite{Perminov2017} or QI.
An ideal broadband QI with an infinite working spectral bandwidth corresponds to $S_0(\nu)=0$, or $F_0(\nu)=1$.
In this case, an exact fulfillment of the equality $F(\nu)=F_0(\nu)$ can be considered as a criterion of the maximum efficiency of the HMR scheme.
Considering the last equality for $S_0(\nu)$ as an approximate condition for the problem of algebraic polynomial optimization, we have:
$F(0)=1,\sum\limits_{m=1}^{N_{opt}}|\operatorname{numer}(F(m\tilde{\Delta}_n/(2N_{opt})))|^2\rightarrow \operatorname{min},N_{opt}=2N-1$
for 13 free parameters $\{1/T_2^*,f_n,g_n,\Delta_n\}$ (where $\operatorname{numer}(F(\nu))$ is the numerator of the rational function $F(\nu)$).
We solved the equality $S_0(\nu)\approx0$ accurately only for several spectral points.
We note that the first of the conditions of the form $F(\nu)=1$ is analogous to the standard condition for a impedance matched resonator QM \cite{Moiseev2010,Moiseev_2017}, which determines the spectral efficiency in the central part of the band for a broadband QM, and additional conditions allow improving the spectral efficiency at edges of the used frequency band.

\begin{figure}[ht]
\begin{center}
\includegraphics[width=0.7\textwidth]{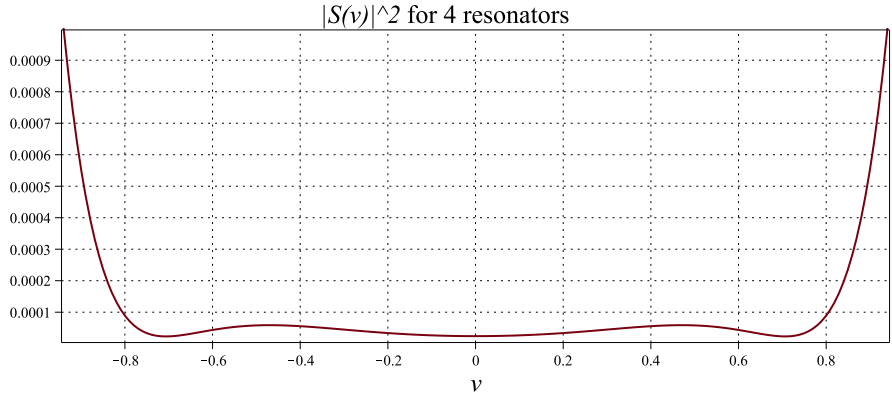}
\end{center}
\caption{
Intensity of the reflected spectrum $|S(\nu)|^2$ for the optimal HMR QI for $\tilde{\gamma}_r=\gamma_n\le10^{-2}$.}
\label{int_load}
\end{figure}

\noindent
After numerical calculations (for $\Delta=1$ and $\tilde{\gamma}_r=\gamma_n=10^{-2}$), we obtain a configuration which is a symmetric with respect to the indices
$1/T_2^*=1.8,f_{\mp1}=1.01,f_{\mp2}=0.707,g_{\mp1}=0.385,g_{\mp2}=0.809,\Delta_{\mp1}=0.56,\Delta_{\mp2}=1.8$ and the intensity for the reflection spectrum $|S(\nu)|^2$ depicted in figure \ref{int_load}.
From where we found the optimal absorption coefficients of the miniresonator modes by spin ensembles $f_1^2T_2^*=0.567$ and $f_2^2T_2^*=0.278$.
The absorption coefficients are much smaller than the optimal absorption coefficient $\kappa$, which should be realized at the usual impedance matching conditions if the spin system will be placed in one common resonator with spectral width $\kappa\gg\Delta=1$ \cite{Moiseev2010}.
This observation shows that the total number of electronic spins can be reduced in the studied QM scheme.
In turn, the realization of the found parameters for the coupling constants $g_1$ and $g_2$ do not cause any problems due to the possibility of significantly enhanced interaction between the mini-resonators and common resonator modes, as it was demonstrated in recent experiment \cite{Moiseev2017_2}.

\noindent
The spectral behavior of $|S(\nu)|^2$ in figure \ref{int_load} demonstrates the quality of the HMR QI for the loss parameters in a common resonator and miniresonators $\tilde{\gamma}_r=\gamma_n\le10^{-2}$.
$|S(\nu)|^2$ is characterized by an almost rectangular spectral plateau of a fairly homogeneous form in the frequency range $\nu\in[-0.8;0.8]$, where the efficiency of the absorption of the signal field by the HMR system reaches $1-|S(\nu)|^2=0.9999$.
We note that with decreasing loss requirements up to $\tilde{\gamma}_r=\gamma_n=10^{-1}$, the absorption efficiency  remains within $99.98\%$ while maintaining the spectral homogeneity of the plateau, which underlines the weak dependence of the optimal configuration from loss parameters.
The transfer efficiency of the signal field to the spin system $\eta_{recording}$ depends to a large extent on the losses during the transfer stage of the signal radiation in the broadband common resonator $\tilde{\gamma}_r$ and it is limited by the characteristic value $e^{-2\tilde{\gamma}_r-2\gamma_n\delta t_s}$ (the upper estimate for the loss in mini-resonators is taken), where $\delta t_s$ is the time duration of the signal emission pulse (in our case $\delta t_s\ge1/\delta \omega_{s,max}=1/(2\cdot0.8\Delta)$, because the size of the plateau in figure \ref{int_load} $\omega_{s,max}=2\cdot0.8\Delta$).
Thus, in order to achieve the transfer efficiency $99.99\%$, the condition $0.95(\tilde{\gamma}_r+\gamma_n/(1.6\Delta))\le10^{-4}$ should be fulfilled which is realistic for the current superconducting technologies.

\section{CONCLUSION}
Summarizing the obtained results, we note that optimization of all the parameters in HMR QI with 4 miniresonators is possible for the signal storage to 4 spin ensembles covering a wide frequency band with a quite large efficiency 99.99\%. The storage requires using sufficiently high-quality miniresonators for which the total losses are limited by $2\tilde{\gamma}_r\sim2\gamma_n\delta t_s\le10^{-4}$.
Rather simple optimization scheme used by us at several spectral points does not require large computational resources and it can be also analyzed analytically on the basis of applied methods of algebraic geometry \cite{Amari1999,Shakirov2007,Dolotin2007,Melnikov2017}.
We note that the HMR scheme allows a very flexible spectral control (due to the dynamic control of parameters \cite{Sandberg2008,Melloni2010,Pierre2014,Wulschner2016,Psychogiou2016,Asfaw2017}) for the transfer of quantum information from an external buffer to a long-lived electron spin ensembles through the system of high-quality miniresonators.
That is, the proposed HMR QI provides an universal efficient scheme for storage of broadband light fields. It can be used to combine several different types of QM devices (for the systems with different types of inhomogeneous broadening) into a single broadband QM block with a higher quality spectral profile \cite{Perminov2017} (see also \cite{Amari2003,Rosenberg2013} about universal spectral blocks and \cite{Amari2006,Swanson2007,Tamiazzo2017} about optimization in the theory of filters).
We think that in the proposed way it is possible to create a superefficient long-lived broadband QM and various hybrid devices \cite{Kurizki2015} involving the use of highly efficient cascaded broadband QI, which can be integrated into the quantum computer circuit on the basis of already existing technologies \cite{Devoret2013,Brecht_2016,Kobe2017,Toth2017} providing the fabrication of a system with strongly coupled high-quality resonators \cite{Megrant2012,Romanenko2014,Huet2016}.

\section*{ACKNOWLEDGMENTS}
This work was partially financially supported by the Russian Foundation for Basic Research through the Grant No. 15-42-02462$\backslash$17 and was partially financially supported by the Grant of fundamental research of the Presidium of the Russian Academy of Sciences 1.26-P.

\bibliography{HMRQM}
\bibliographystyle{spiebib}

\end{document}